\documentclass{jfm}
\usepackage{graphicx}
\usepackage{epstopdf, epsfig}
\usepackage{amsmath}

\newcommand\Ray{\mbox{\textit{Ra}}}
\newcommand\Nus{\mbox{\textit{Nu}}}
\newcommand\Ric{\mbox{\textit{Ri}}}

\title{Transition in the asymptotic suction boundary layer over a heated plate}
\author{S. Zammert \aff{1,2}
  \corresp{\email{Stefan.Zammert@gmail.com}},
  N. Fischer\aff{3}
 \and B. Eckhardt\aff{1,4}}

\shorttitle{Asymptotic suction boundary layer over a heated plate}
\shortauthor{S. Zammert, N. Fischer and B. Eckhardt}

\affiliation{\aff{1} Fachbereich Physik, Philipps-Universit\"at Marburg, D-35032 Marburg, Germany\\[\affilskip]
\aff{2} Laboratory for Aero and Hydrodynamics, Delft University of Technology,\\ 2628 CD Delft, The Netherlands\\[\affilskip]
\aff{3}Fakult\"{a}t Verkehrswissenschaften ``Friedrich List'', Technische Universit\"at Dresden, D-01062 Dresden, Germany\\[\affilskip]
\aff{4}J.M. Burgerscentrum, Delft University of Technology, 2628 CD Delft, The Netherlands}

\pubyear{2012}
\volume{650}
\pagerange{119--126}
\begin{document}

\maketitle

\begin{abstract}
%The asymptotic suction boundary layer (ASBL) is a parallel shear flow that becomes turbulent
%in a bypass transition in parameter regions where the laminar profile is stable. 
%We here add a temperature gradient perpendicular
%to the plate and explore the interaction between convection and shear in determining the transition.
%We find that the laminar state becomes unstable in a subcritical bifurcation and that
%the critical Rayleigh number and wave number depend strongly on the Prandtl number. We also track
%several secondary bifurcations and identify states that are localized in two directions, 
%showing different symmetries.
%In the subcritical regime, transient turbulent
%states which are connected to exact coherent states and follow the same transition scenario
%as found in linearly stable shear flows are identified and analyzed. 
%The study extends the bypass transition scenario from shear flows to 
%thermal boundary layers and shows the intricate interactions between thermal and shear forces in 
%determining critical points.

The asymptotic suction boundary layer (ASBL) is a parallel shear flow that becomes turbulent
in a bypass transition in parameter regions where the laminar profile is stable. 
We here add a temperature gradient perpendicular
to the plate and explore the interaction between convection and shear in determining the transition.
We find that the laminar state becomes unstable in a subcritical bifurcation and that
the critical Rayleigh number and wavenumber depend strongly on the Prandtl number. We also track
several secondary bifurcations and identify states that are localized in two directions, 
showing different symmetries.
In the subcritical regime, transient turbulent
states which are connected to exact coherent states and follow the same transition scenario
as found in linearly stable shear flows are identified and analyzed. 
The study extends the bypass transition scenario from shear flows to 
thermal boundary layers and highlights the intricate interactions between thermal and shear forces. 
%in  determining critical points.

\end{abstract}
 
\begin{keywords}
%Authors should not enter keywords on the manuscript, as these must be chosen by the author during the online submission process and will then be added during the typesetting process (see http://journals.cambridge.org/data/\linebreak[3]relatedlink/jfm-\linebreak[3]keywords.pdf for the full list)
\end{keywords}

\section{Introduction}

Thermal transport in shear flows is a fundamental problem in 
fluid mechanics that arises in many engineering and geophysical 
situations. It combines phenomena in fluid layers without
external shear (the Rayleigh-B\'enard (RB) flow, \citet{Chandrasekhar1961}) and in shear
flows without heating \citep{Grossmann2000,Henningson}. The onset of convection in the RB setting
has been studied in considerable detail in various geometries
%, where the fact that they can be done in closed domains simplifies the analysis 
\citep{Busse1978}. 
The properties of shear flows are less well understood because of the
difficulties presented by the manner in which they become turbulent \citep{Grossmann2000,Henningson}. 
In contrast to RB flows, turbulence in shear flows is often observed
when the laminar profile is still linearly stable, so that the transition
requires with finite amplitude perturbations that reach states that
arise in subcritical bifurcations \citep{Eckhardt2007c}. The advective nature and the spatial development in open
boundary layers add to the complexity.

The interaction between shear and convection has been studied in boundary layers \citep{Sparrow1961,Merkin1972},
where it affects the functional form of the profile \citep{Marati2006}. The dynamics of plumes and other
structures %in the boundary layer 
has been investigated in the context of RB flows \citep{Ahlers2009,Chilla2012} 
because of their contribution to the heat transport \citep{Zocchi1990,Parodi2004,Zhou2007,Zhou2010c}. 
The presence of instabilities is connected with  a transition that also
marks the entry into the ultimate regime \citep{Grossmann2000a,Grossmann2002}. Experiments by  \citet{DuPuits2013} 
show that the flow in the transition region
can be spatially localized and transient, very much as in the case of the bypass transition
in shear flows. A  quantitative study focusing on the hydrodynamic instabilities in heated 
boundary layers does not seem to have been published.

The manner in which shear and buoyancy interact to shape the transition
has been studied by \citet{Clever1992} for the case of RB with a linear shear.
They identified the onset of  modulated rolls and the transition to a subcritical regime when the
shear was strong and the temperature gradient weak. Ultimately, this state
could be traced to an exact coherent structure (ECS) in plane Couette flow
that is related to the transition in the pure shear flow without a temperature gradient. 
The ECS reached via heating coincides with the one obtained by adding rotation
as a destabilizing force \citep{Nagata1990}. Using these and related embeddings as well as some other techniques \citep{Dijkstra2013}, numerous exact coherent structures have been identified in plane Couette flow \citep{Nagata1990,Schmiegel1999,Gibson2009}, in pipe flow 
\citep{Faisst2003,Wedin2004,Hof2004,Pringle2007,Eckhardt2008c,Duguet2008}, in plane Poiseuille flow  \citep{Waleffe2001,Nagata2013a,Gibson2014,Zammert2014a,Zammert2014b} and even in boundary
layers \citep{Duguet2012}. For these flows the observed transient subcritical turbulence could
be explained by ECS which create a chaotic saddle in a series of bifurcations \citep{Kreilos2012,Avila2013,Zammert2015}. 
Further studies have shown that ECS are linked dynamically \citep{Halcrow2009} and embedded 
in the turbulent dynamics \citep{Kawahara2001}.  

The asymptotic suction boundary layer (ASBL) \citep{Schlichting1997} is a boundary layer that is of interest
on its own because of its relation to flow control since suction stabilizes the boundary layer. It is also of 
theoretical interest since the flow is parallel and can hence be studied using the tools developed for parallel
shear flows. Experimental realizations of both the pure shear flow \citep{Fransson2003} and the heated version
\citep{Moffat1968} are available. Taking the cross flow as a parameter, the ECS from plane Couette flow could be traced to
ASBL, and a transition in the dynamics could be identified \citep{Kreilos2013,Kreilos2016}. Further ECS for ASBL, 
in the form of  travelling waves and periodic orbits, have been identified and studied \citep{Khapko2013,Khapko2013a,Deguchi2014c}.
We here expand the analysis of the ASBL to include heating of the bottom plate, so as to form a thermal asymptotic suction boundary layer (TASBL).
The study of such a parallel TASBL contributes to our understanding of the turbulence transition and the 
dynamics of coherent structures in thermal boundary layers. 

The further outline of this paper is as follows: in section 2, we introduce the system and its laminar solution. The stability of the
base flow as function of Rayleigh number, Reynolds number, and Prandtl number 
and the two-dimensional secondary solutions bifurcating from the laminar state are discussed in section 3. 
In section \ref{secSubChaos} examples of the cascade of bifurcations that lead to a chaotic saddle and subcritical chaos
in the system are studied. Finally, in section \ref{sec3Dstates} three-dimensional secondary bifurcations are analysed,
with a particular emphasis on exact solutions that are localized in both directions parallel to the plate. We end
with an outlook in section \ref{SecConcl}.

\section{Description of the system and its laminar state}
\begin{figure}
\centering
\includegraphics[]{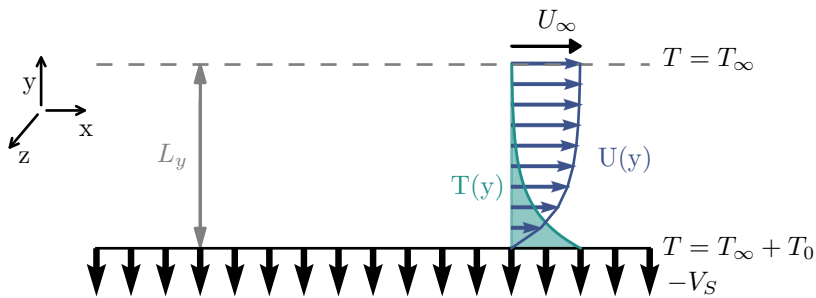}
\caption{Sketch of the ASBL over a heated plate.
Far above the plate the flow moves with speed $U_{\infty}$ in streamwise direction. The temperature difference between 
the plate and the fluid far above the plate is $T_{0}$. 
In the numerical simulation the domain is closed from above with a plate at a height $L_{y}$, indicated by 
a gray line. \label{fig_System}}
\end{figure}

The asymptotic suction boundary layer is the flow over a plate through which fluid is
sucked homogeneously with a speed $-V_{S}$. Far above the plate
the parallel velocity component of the flow has a speed $U_{\infty}$.
For TASBL there is a temperature difference $T_{0}$ between the plate and the fluid far above the plate.
The flow geometry of TASBL is sketched in figure \ref{fig_System}.

The flow is governed by the Overbeck-Boussinesq equations \citep{Landau1959}
\begin{equation}
 \partial_{t} \vec{u} + (\vec{u}\cdot \nabla)\vec{u} = -\nabla p + \nu \Delta \vec{u} + \beta g \theta \vec{e}_{y} 
 \label{OBeq1}
\end{equation}
\begin{equation}
 \partial_{t} \theta + (\vec{u}\cdot \nabla)\theta = \kappa \Delta \theta 
 \label{OBeq2}
\end{equation}
\begin{equation}
 \Delta \vec{u} = 0 
 \label{OBeq3} .
\end{equation}
where $\nu $ is the kinematic viscosity, $\kappa$ the thermal diffusivity, $\beta$ the thermal expansion coefficient
and $g$ the gravitational acceleration, pointing in the $y$-direction perpendicular to the plate. 
The boundary conditions of the system are
\begin{equation}
 u(y=0)=0 \ \text{and} \ u(y=\infty)=U_{\infty}
\end{equation}
\begin{equation}
 v(y=0)= v(y=\infty)=-V_{S}
\end{equation}
\begin {equation}
 \theta(y=0)=T_{0} \ \text{and} \ \theta(y=\infty)=0 .
\end {equation}
The stationary laminar profiles for the velocity is given by
\begin{equation}
 \vec{U}(x,y,z)=(U(y),-V_{s},0)  \ \text{with} \  U(y)=U_{\infty}(1-e^{-V_{s}y/\nu})
\end{equation}
and the deviation of the temperature from the background temperature $T_\infty$ is given by
\begin{equation}
 T(x,y,z)=T_{0}\cdot e^{-V_{s} y /\kappa}.
\end{equation}
A characteristic length scale of the system is the displacement thickness \citep{Schlichting1997}
\begin{equation}
 \delta=\int_{0}^{\infty} \left(1- \frac{U(y)}{U_{\infty}} \right) dy=\frac{\nu}{V_{S}} .
\end{equation}
The dimensionless numbers characterizing the system are the Reynolds, Prandtl and Rayleigh number, given by 
\begin{equation}
\Rey=\frac{U_{\infty}\delta}{\nu}=\frac{U_{\infty}}{V_{s}}, \ \ \ \Pran =\frac{\nu}{\kappa}, \ \ \ \Ray=\frac{\beta g\delta^{3}T_{0}}{\kappa \nu}. \label{eqnRaRePrDefs}
\end{equation}

The equations can be made dimensionless by measuring lengths in units of $\delta$, time in units of $\delta^{2} / \kappa$ (corresponding to measuring velocities in units of $\kappa / \delta$) and temperature
in units of $T_{0}/\Ray$:
\begin{equation}
\partial_{t} \vec{u} + (\vec{u}\cdot \nabla)\vec{u} = -\nabla p + Pr \Delta \vec{u} + \Pran \Ray \ \theta  \vec{e}_{y} \label{NSeqn1}
\end{equation}
\begin{equation}
\partial_{t} \theta + (\vec{u}\cdot \nabla)\theta = \Delta \theta \label{NSeqn2}
\end{equation}
\begin{equation}
\nabla \cdot \vec{u} =0  \label{NSeqn3}
\end{equation}
The non-dimensionalized laminar profiles become
\begin{equation}
\vec{U}(x,y,z)=(U(y),-Pr,0) \  \ \  U(y)=\Pran \Rey (1-e^{-y})
\end{equation}
\begin{equation}
T(x,y,z)= \Ray \ e^{-Pr \ y}
\end{equation}

The thickness $\delta_{\theta}$ of the thermal boundary layer is then
inversely proportional to the Prandtl number,
\begin{equation}
\delta_{\theta}=\frac{\delta}{\Pran}.
\end{equation}  
Thus, the Prandtl number can also be interpreted as the ratio between the thicknesses of the velocity boundary layer $\delta$ 
and the thermal boundary layer $\delta_{\theta}$.
For values of $\Pran$ smaller than unity (e.g air and liquid metals) the thermal boundary
layer is thicker than the velocity boundary layer. For $\Pran=1$ both have the same thickness, while for $\Pran>1$ 
(e.g. water and oils) the thermal boundary layer is the thinner one.

The  velocity and temperature fields are decomposed into a base flow and fluctuations,
\begin{equation}\theta=T + \theta',\end{equation}
\begin{equation}\vec{u}=\vec{U} + \vec{u}'.\end{equation}
Unless stated explicitly, only the fluctuations $ \theta'$ and $\vec{u}'$ are visualized or used in bifurcation diagrams.

%Truncation in the simulations
In the numerical simulations the wall-normal direction has to be truncated at some distance $L_{y}$.
To this end, the system is closed with a permeable plate with asymptotic temperature and velocity 
components, and no-slip boundary conditions for the deviations.
The modified base profiles become
\begin{equation}
 U^{*}(y)=\Rey \Pran\frac{1-e^{-y}}{1-e^{-L_{y}}} 
\end{equation}
and
\begin{equation}
 T^{*}(y)=\frac{\Ray}{1-e^{-Pr \ L_{y}}}\left( e^{-Pr \ y } -e^{-Pr \ L_{y}} \right) .
 \end{equation}
In all simulations a value of $L_{y}$ equal to or larger than $20$ is used. In all simulations it was carefully checked
that this value is sufficiently big so that the differences between the asymptotic profiles $U(y)$ and  $T(y)$ and 
the finite profiles  $U^{*}(y)$ and $T^{*}(y)$ remained small.

%%Nusselt number
A quantity of particular interest in the study of thermal systems is the heat flux,
which has a convective and a conductive contribution. For the base flow,
the convective heat flux is given by 
\begin{equation}
 \phi_{conv}(y)=  \left< v\cdot  T \right>_{x,z}=-Pr Ra \ e^{-Pr \ y }.
\end{equation}
For the conductive heat flux we find
\begin{equation}
 \phi_{cond}(y)= - \partial_{y} \left< T \right>_{x,z}=Pr Ra \ e^{-Pr \ y }.
\end{equation}
Thus, in contrast to Rayleigh-B\'{e}nard convection, the conductive heat flux of the base flow depends on the wall-normal coordinate $y$.
Moreover, the conductive and convective contributions cancel, and the total heat transfer of the laminar state vanishes.
This is no longer the case for the turbulent states. 
As a dimensionless measure for the heat flux we introduce the Nusselt number $Nu$, 
the ratio of the total heat flux to the conductive heat flux of
the base flow at the wall $\phi_{cond}(0)$,
\begin{eqnarray}
 Nu&=&\frac{\phi_{conv} +\phi_{cond}}{\phi_{cond}(0)} \\
   &=&\frac{\left< (-Pr + v)( \theta'+T)\right>_{x,z} - \partial_{y}\left<\theta' + T \right>_{x,z}}{Pr Ra}. \label{DefNu1}
\end{eqnarray}
In order to eliminate the dependence on the wall-normal coordinate, we integrate
over $y$ and introduce
\begin{equation}
 \tilde{\Nus}=\int_{0}^{\infty} \Nus(y) \ dy  , \label{DefNu2}
\end{equation}
the volume-integrated heat flux.

\section{Stability analysis and secondary solutions}
In this section the stability of the laminar state is analysed and the bifurcating secondary solutions are discussed.
The linear stability of the flow is determined by solving the Orr-Sommerfeld equations for the flow.
The derivation of the equations follows the usual steps and is described in subsection \ref{Section_OSEqn}. 
Direct numerical simulation is used to identify exact solutions of the equations. For the simulations
we used the \textit{Channelflow}-code \citep{J.F.Gibson2012} (www.channelflow.org), which we enhanced to handle a temperature field.
Details of the modification are given in appendix \ref{AppendixDNS}.

Following the derivation of the stability equations, we will first analyse the case $\Ray=0$ in subsection  \ref{Sec_StabilityASBL_Ra0}.
Afterwards, we will present the result for the case $\Rey=0$ in subsection \ref{sec_StabRe0}. Finally, the combined effect of temperature and shear
is discussed in subsection \ref{sec_StabReVar}.

\subsection{Orr-Sommerfeld stability equations\label{Section_OSEqn}} 
Starting point for the stability analysis are the Navier-Stokes equations (\ref{NSeqn1}-\ref{NSeqn3}).
The full flow fields $\vec{u}$ can be expressed as a sum of a base flow contribution $\vec{U}$ and a perturbation $\vec{u}'$.
The temperature and pressure fields can be decomposed similarly.
Inserting this ansatz in the equations and neglecting the non-linear terms in the perturbations we obtain:
\begin{equation}
 \frac{1}{\Pran} \left[\partial_{t} \vec{u}' + 
 %(\vec{u}'\cdot \nabla)\vec{u}' +
 % Die nichtlinearen Terme sollten doch raus, oder?
 (\vec{u}'\cdot \nabla)\vec{U}+(\vec{U}\cdot \nabla)\vec{u}'\right]= 
 -\nabla p + \Delta \vec{u}' +  \Ray (T+\theta') \vec{e}_{y}
 \end{equation}
\begin{equation}
\partial_{t} \theta' + (\vec{u}'\cdot \nabla)T + \vec{U}\cdot \nabla \theta' = \Delta \theta' 
 \end{equation}
Form this set of equations the pressure can be removed by writing the  solenoidal perturbation velocity field $\vec{u}'$ in terms of two scalar fields $\Phi(\vec{x})$ and $\Psi(\vec{x})$.
\begin{equation}
\vec{u}'=\nabla \times \Psi \vec{e}_{y} + \nabla \times (\nabla \times \Phi \vec{e}_{y})
\end{equation}
For the resulting three scalar fields $\Phi$, $\Psi$ and $\theta'$ we take Fourier series
in spanwise and streamwise direction and an exponential dependence in time,
\begin{equation}
\Phi(\vec{x},t) = \sum_{k_{x},k_{z},\sigma}{ \Phi_{k_{x},k_{z},\sigma}(y)\exp{(i(k_{x}x+k_{z}z)+\sigma t})}
 \end{equation}
 \begin{equation}
\Psi(\vec{x},t) = \sum_{k_{x},k_{z},\sigma}{ \Psi_{k_{x},k_{z},\sigma}(y)\exp{(i(k_{x}x+k_{z}z)+\sigma t})}
 \end{equation}
 \begin{equation}
\theta'(\vec{x},t) = \sum_{k_{x},k_{z},\sigma}{ \Theta_{k_{x},k_{z},\sigma}(y)\exp{(i(k_{x}x+k_{z}z)+\sigma t})}
 \end{equation}
Using this representation of the fields, we obtain the stability equations
\begin{equation}
\left( \frac{\sigma}{Pr} + \frac{ik_{x}U}{Pr} -  D_{y} - (D_{y}^{2} -k^{2})\right)\Psi(y) + \frac{ik_{z}U'}{Pr}\Phi(y)=0 \label{OSeqn1}
\end{equation}
\begin{equation}
\left[\left(\frac{\sigma}{Pr}+\frac{ik_{x}U}{Pr}-D_{y}\right)(D_{y}^{2}-k^{2}) - \frac{ik_{x}U''}{Pr} - (D_{y}^{2} -k^{2})^{2}\right]\Phi(y) +Ra\Theta(y)=0 \label{OSeqn2}
\end{equation}
\begin{equation}
\left(\sigma  + ik_{x}U - Pr D_{y} - (D_{y}^{2} - k^{2}) \right)\Theta(y) + k^{2}T' \Phi(y)=0 \label{OSeqn3}
\end{equation}
In this set of equations $D_{y}$ denotes the derivative with respect to the wall-normal coordinate, % $y$,
$U(y)=\Rey \Pran (1-e^{-y})$ is the streamwise component of the laminar velocity field and 
$T(y)=e^{-Pr y}$ is the laminar temperature profile. 
$k_{x}$ and $k_{z}$ are the streamwise and spanwise wave numbers, and $k=\sqrt{k_{x}^{2} +k_{z}^{2}}$ is the wave number in the direction of propagation.
The set of equations \ref{OSeqn1}-\ref{OSeqn3} is solved using a Chebychev spectral method \citep{Trefethen2000},
where the Chebychev polynomials are mapped from an interval $[-1,1]$ to an interval $[0,L_{y}]$ with a sufficiently
large truncation length $L_{y}$. % has to be chosen sufficiently big.

\subsection{Stability analysis for \Ray=0 \label{Sec_StabilityASBL_Ra0}}
\begin{figure}
\centering
\includegraphics[]{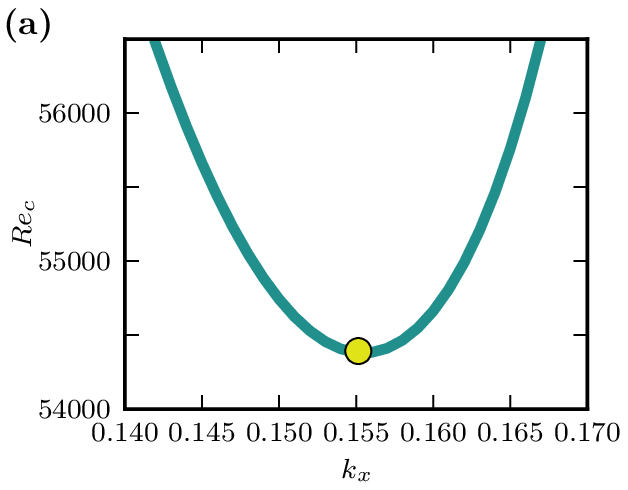}\includegraphics[]{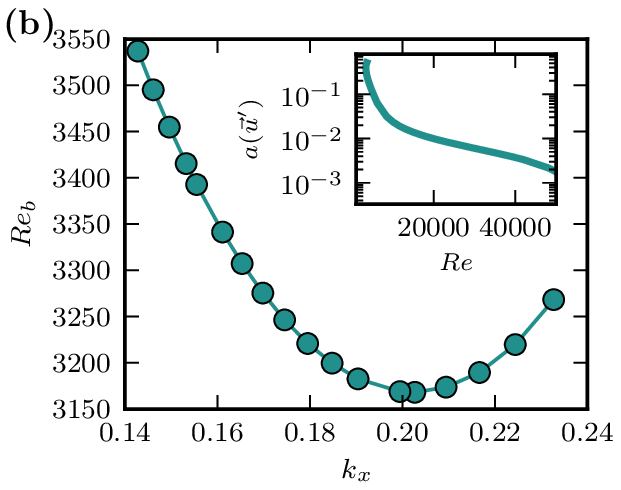}
\caption{The stability boundary of the laminar state of isothermal ASBL (\Ray=0) is shown in panel (a). 
The critical Reynolds number $Re_{c}=54379$ is marked by a circle. Panel (b)  shows the Reynolds number $Re_{b}$ of the 
subcritical bifurcation point 
%of the bifurcating 
for the TS-wave vs. streamwise wave number $k_{x}$.
The inset shows a bifurcation diagram of the TS-wave for $k_{x}=k_{c}$ which uses the amplitude of the flow field
$a(\vec{u}')$ on the vertical axis.%=\sqrt{(1/V \int_{V} \vec{u}'^{2} dV)}$ 
\label{fig_StabDiagRa0}}
\end{figure}

For $\Ray=0$ the system reduces to isothermal ASBL.  \cite{Hocking1975a} showed that the laminar profile of isothermal ASBL has
a linear instability at \Rey $\approx 54370$ for a critical streamwise wave number $k_{c}=0.1555$.
The fact that ASBL becomes unstable at such a  high Reynolds number compared to the Blasius boundary layer which becomes unstable at $Re_{c,BL}=519.4$ \citep{Henningson} shows 
the strongly stabilizing effect of the suction. \cite{Milinazzo1985} 
tracked the travelling wave solutions  which bifurcate from the laminar profile and
found that they extend down to Reynolds numbers around 3000. 
To obtain the stability curve the stability equations are solved with up to 
$500$ Chebychev polynomials and a truncation length of $40$ in $y$-direction.
Using a bisection the  critical value for $\Rey$ is reproduced with a relative error of order $10^{-2}$. 
Squires theorem  \citep{Gage68,Henningson}
ensures that a two-dimensional streamwise mode is the first to become unstable so that we can restrict
the Orr-Sommerfeld analysis to spanwise invariant modes ($k_{z}=0$).
The critical Reynolds numbers as a function of streamwise wavenumber
agree with the results of \cite{Hocking1975a} and the more recent ones of \cite{Fransson2003};
they are shown in figure \ref{fig_StabDiagRa0}(a). The critical Reynolds number found here is $54379$,
which is achieved for a streamwise wave number $0.1555$.
\cite{Wedin2015} showed recently that the instability of the flow moves to lower 
Reynolds numbers if the plate's permeability is not neglected.

The Tollmien-Schlichting-like travelling wave solution which bifurcates subcritically from the laminar flow 
%at the point of the linear instability 
has been traced in $\Rey$ for various values of the streamwise wave number.  
A bifurcation diagram of the waves is included in figure \ref{fig_StabDiagRa0}(b) in an inset. 
For the ordinate of the diagram we use the amplitude,
\begin{equation}
a(\vec{u}')=\sqrt{\left(\frac{1}{L_{x}L_{z}} \int_{L_{x}}\int_{L_{y}}\int_{L_{z}}  \vec{u}'^{2} \ dx \ dy \ dz\right)},
\end{equation}
of the flow field, where $L_{x}$ and $L_{z}$ are the streamwise and spanwise wavelengths of the computational domain.
The plot shows that close to the turning point the amplitude of the flow field increases dramatically.
The dependence of the positions of the turning point on the wave number is given in figure \ref{fig_StabDiagRa0}(b). The minimal Reynolds number of the turning point is $3168$, which is achieved for $k_{x}=0.203$.
These results show that as in the case of plane Poiseuille flow the two-dimensional waves connected to the 
instability of the base flow appear at Reynolds numbers much higher than the observed onset of subcritical
turbulence, which was found to be at $Re\approx 270$ for ASBL \citep{Khapko2016}.

\begin{figure}
\centering
\includegraphics[]{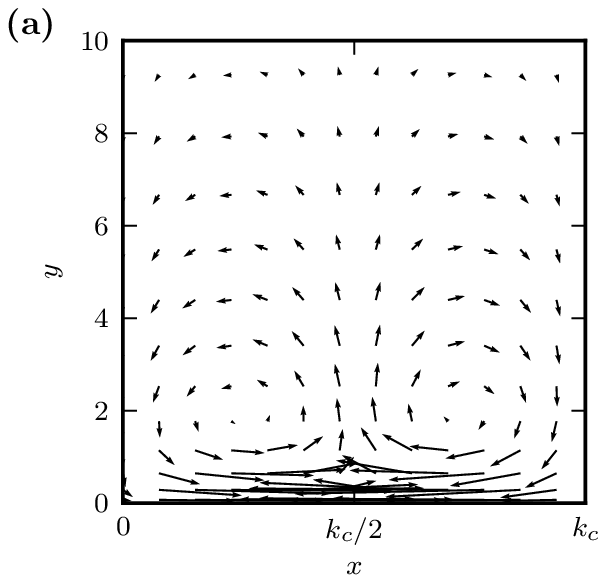}\includegraphics[]{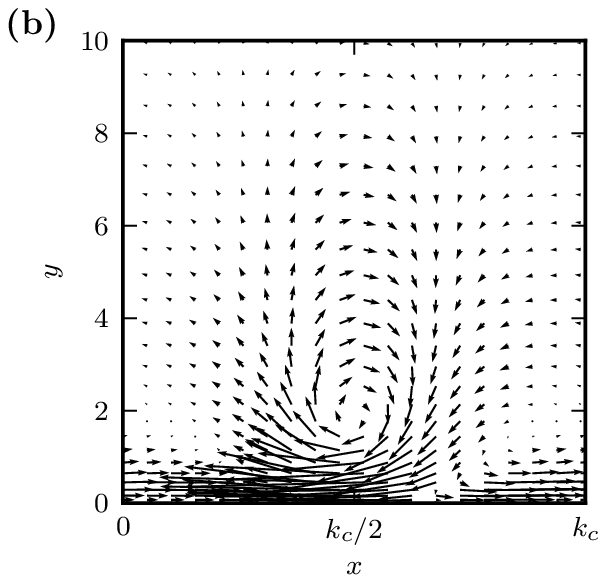}
\caption{Visualizations of the TS-wave bifurcating from the base flow for $Re=50000$ (a) and $Re=8475$ (b). In both cases the streamwise wave number is $0.1555$. The direction of the flow is from left to right.\label{fig_TSWavesASBL}}
\end{figure}

Visualizations of the bifurcating two-dimensional solutions are given in figure \ref{fig_TSWavesASBL} for different Reynolds numbers.
Because the flow has no discrete symmetry with respect to the wall-normal direction $y$, the clockwise and counter-clockwise rolls
behave differently with $\Rey$.  Close to the bifurcation point at $Re=50000$ (figure \ref{fig_TSWavesASBL}a) the rolls are almost symmetric. 
When $Re$ is decreased the counter-clockwise roll becomes strongly deformed while the clockwise roll maintains its 
shape and becomes narrower only (figure \ref{fig_TSWavesASBL}b). 

\subsection{Stability Analysis for \Rey=0 and $\Ray\neq0$\label{sec_StabRe0}}
\begin{figure}
\centering
\includegraphics{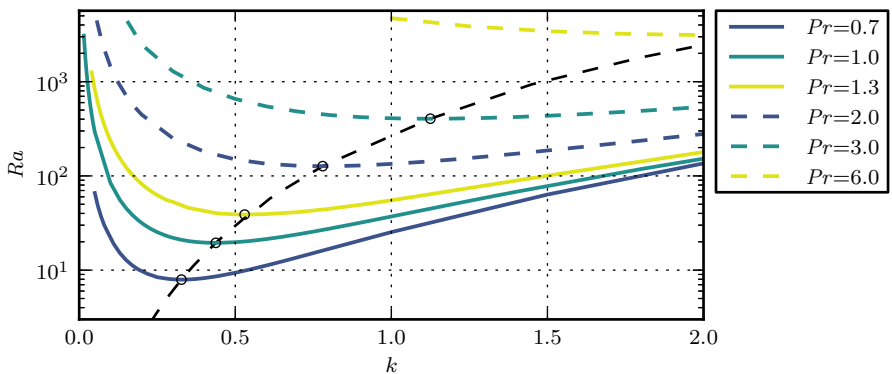}
\caption{Stability boundary for \Rey=0  and different values of the Prandtl number. For each Prandtl number
the lowest critical Rayleigh number is marked by a black circle. 
\label{fig_StabDiagPrVar}}
\end{figure}

We now turn to the stability of the thermal boundary layer for the case \Rey=0.  The corresponding Orr-Sommerfeld equations (equation \ref{OSeqn1}-\ref{OSeqn3}) are solved using
up to $600$ Chebychev polynomials and truncation lengths $L_{y}$ up to $250$ (for small Prandtl numbers).
In figure \ref{fig_StabDiagPrVar} the calculated stability curves for various Prandtl numbers, including the 
cases \Pran=0.7 (air) and \Pran=6 (water), are shown.  
The results show a strong dependence on the Prandtl number and therefore on the relative size of the thermal to the viscous boundary layer
that is in contrast to the standard Rayleigh-B\'enard convection. For the case $\Pran =1$ a critical Rayleigh 
number of  $19.4705$ and a critical wave number of $0.438$ are obtained. 
For lower values of $\Pran$ the critical wave number and Rayleigh number move to lower values, while they
move  to higher ones if larger values of $\Pran$ are considered.

\begin{figure}
\centering
\includegraphics[]{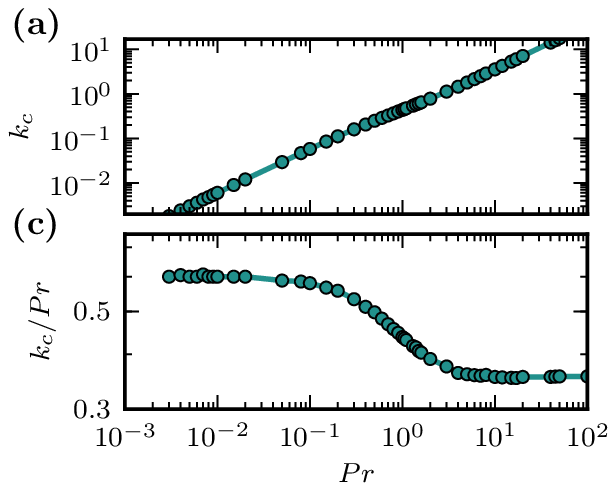} 
\includegraphics[]{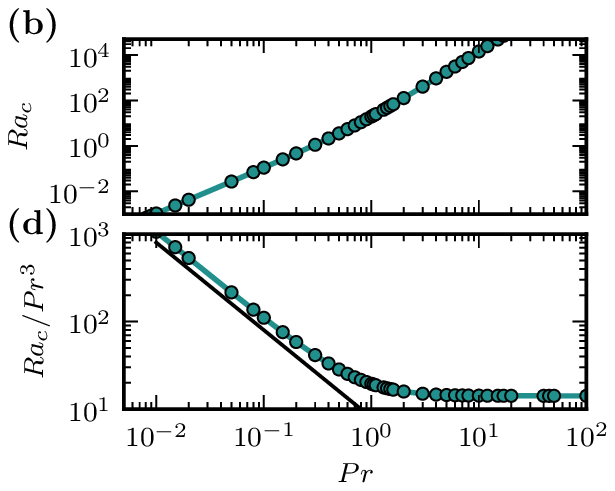}
\caption{Prandtl number dependence of critical wave number and critical Rayleigh number as a function of Prandtl number
are shown in panels (a) and (b), respectively.
Panels (c) and (d) show the results after a change of the length scale to the thermal boundary layer thickness 
$\delta_{\theta}=\delta/\Pran$.
The thin black line in (d) indicates a $\Pran^{-1}$ scaling. % is shown in (d). 
\label{fig_kcritVsPr}}
\end{figure}

The critical Rayleigh and wave number are shown in figure \ref{fig_kcritVsPr}(a) and (b) in dependence on the Prandtl number. 
The graphs show that the dependence of the critical wave number on $\Pran$ is approximately linear.
The $\Pran$-dependence of the critical Rayleigh number seems to follow a power law,
with an exponent near two for  $\Pran<1$ and near three for $\Pran>1$.

An explanation for the $\Pran$-dependence of the critical Rayleigh number may be obtained from a discussion
of the relevant scales, similar to the one given by \cite{Grossmann2001} for normal Rayleigh-B\'enard convection. 
The definition of $\Ray$ uses the length scale of the velocity boundary layer $\delta$ 
and the full temperature difference. 
For large values of $\Pran$ the thickness of the temperature boundary layer $\delta_{\theta}$ is much smaller than the one of the velocities.
Therefore, the length scale relevant for the forces driving convection is $\delta_{\theta}$.   A Rayleigh number based on $\delta_{\theta}$ 
can be obtained by multiplication of $\Ray$ with $1 /\Pran^{3}$ 
because $\Ray \propto \delta^{3}$ and $\delta / \delta_{\theta} = \Pran$. By the same argument, the wave number based on $\delta_{\theta}$ 
is obtained from the one based on $\delta$ by dividing by $\Pran$.
The results of this change in length scale are shown in figure \ref{fig_kcritVsPr}(c) and (d).
For the wave number as well as for $\Ray$ the change of the length scale leads to critical values that for large values of $\Pran$ become independent of the Prandtl number.
The critical wave number $k_{\theta,c}$ and Rayleigh number $\Ray_{\theta}$ based on the thickness of the temperature boundary layer that
are asymptotically reached for large Prandtl numbers are $0.363$ and $14.3$, respectively.

For the case $\Pran < 1$, two modifications have to be taken into account. 
The observation that the rescaled critical wave number reaches a plateau value for small values of $\Pran$ (see figure \ref{fig_kcritVsPr}(c)), 
suggest that the thickness of the thermal boundary layer determines the instability also for this parameter regime.
However, since the viscous boundary layer is smaller than the thermal boundary layer, the temperature drops only by
a fraction $(\delta / \delta_{\theta}) T_{0}=\Pran  T_{0}$, and it is this smaller range that should be taken for the
temperature difference in the definition of the Rayleigh number. Accordingly, there is a factor of $\Pran$ from the
temperature difference, and a factor $\Pran^{-3}$ from the scaling of the height, resulting in a scaling of $\Pran^2$
for the critical Rayleigh number,  or $\Pran^{-1}$ in the scaling used in figure  \ref{fig_kcritVsPr}(d).

By using the Newton method \citep{Viswanath2007} included in the channelflow package, 
it is possible to identify the exact solutions bifurcating at the linear instability and to continue them in Rayleigh number. 
The used resolution is $N_{x}\times N_{y} \times N_{z}=4\times 49 \times 32$. 
Four modes in the $x$-direction are sufficient because the structures are two-dimensional and do not vary in $x$-direction. In
wall-normal direction a truncation length $L_{y}$ is chosen which is large enough so that there is no 
noticeable influence on the
structures by the wall. For lower branch states $L_{y}=40$ is sufficient, while for upper branch states values up to $L_{y}=110$ with a resolution $N_{y}=129$ are necessary. 

\begin{figure}
\centering
\includegraphics[]{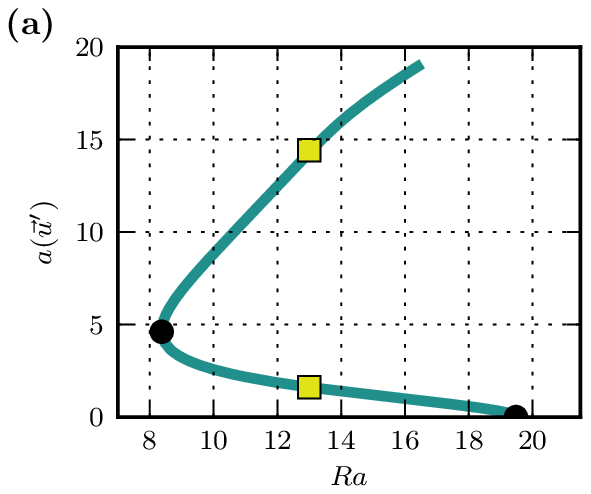}\includegraphics[]{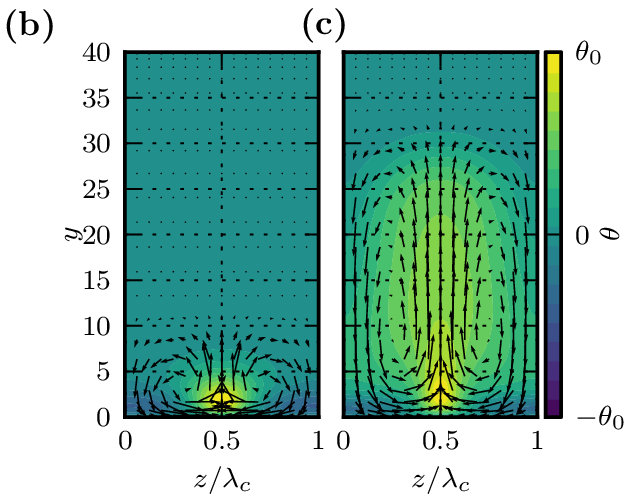}
\caption{Panel (a) shows the bifurcation diagram of the secondary solution with $k=k_{c}=0.4305$. On the $y$-axis the amplitude
$a(\vec{u}')$ of the solution is used. Panels (b) and (c) show visualizations of the 
flow- and temperature fields for the lower and the upper branch of $R_{1}$. The positions of the visualization are marked in (a) using yellow squares.
The velocity field is visualized by the arrows. The temperature field is represented by the colours. 
For the lower branch $\theta_{0}=0.19\Ray$ and for the upper branch $\theta_{0}= 0.32\Ray$. \label{fig_FlowFieldsRa13}}
\end{figure}

The bifurcating ECS identified by the Newton method is an  equilibrium solution which we  refer to as roll state $R_{1}$ in the following.
A bifurcation diagram of the ECS for the critical wave number $k_{c}$ and $\Pran=1$ is given in figure \ref{fig_FlowFieldsRa13}(a). 
The diagram shows that in contrast to the case of Rayleigh-B\'enard flow the bifurcation is subcritical.
For the critical wave number $k_{c}=0.4304$ used for the bifurcation diagram, the turning point of the solutions
lies at $\Ray =8.374$ but this value varies with wave number. With decreasing wave number the turning point moves to lower values of $\Ray$ while with increasing wave number it moves to higher Rayleigh numbers.  

\begin{figure}
\centering
\includegraphics[]{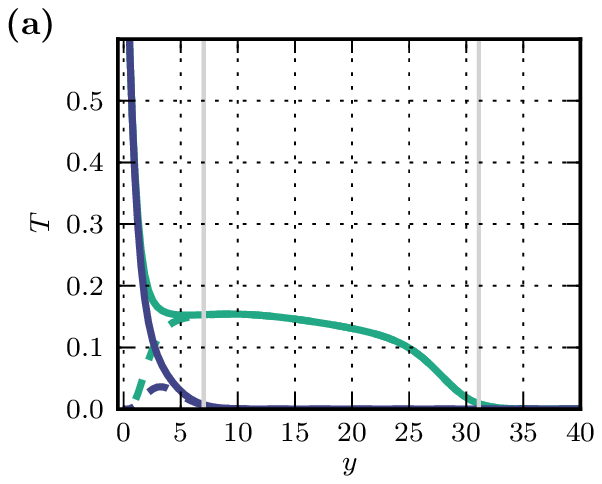}
\includegraphics[]{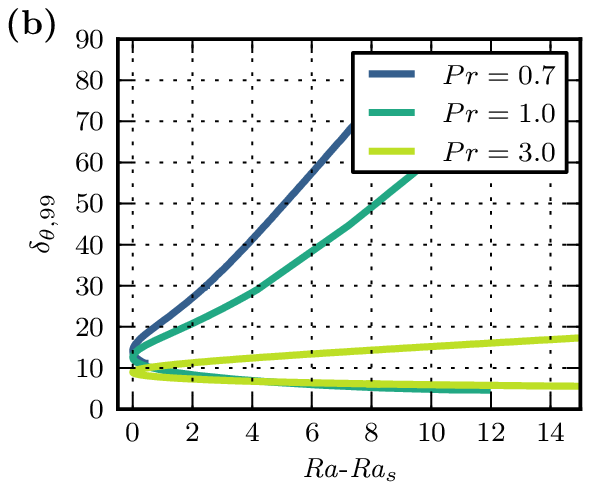} 
\caption{(a) Temperature profiles of the lower (blue) and upper branch (cyan) of the
secondary solution $R_{1}$ at $\Ray=13$ %in dependence on wall-normal coordinate $y$
vs. distance $y$ above the plate. 
For the total temperature profile a solid line is used and the
deviations from the base profile are shown by dashed lines. The positions of $\delta_{\theta,99}$ are marked using gray lines.
(b) Values of $\delta_{\theta,99}$ of the secondary solution $R_{1}$ in dependence on the distance to the saddle-node point  for different Prandtl numbers.
 For each Prandtl number the wavelength of the state shown is chosen to be %tthe secondary solution with 
%that for 
the critical wavelength at the particular value of the Prandtl number.
 \label{fig_HeightVsRa}}
\end{figure}

The flow and temperature fields for the lower and the upper branch of the state shown in figure \ref{fig_FlowFieldsRa13}(b) and (c) reveal 
that the secondary solution $R_{1}$ consists of counter-rotating convection rolls. With increasing distance from the bifurcation
point rolls grow in height while maintaining their width, so that 
their aspect ration (height$/$width) increases.
To quantify the growth in height we use the thickness $\delta_{\theta,99}$, %which is 
defined as the distance to the plate at which %where 
the mean temperature profile
reaches $0.01T_{0}$ for the first time. The definition of $\delta_{\theta,99}$ is illustrated  in figure \ref{fig_HeightVsRa}(a) using
temperature profiles of both branches of $R_{1}$.
The variation of the quantity $\delta_{\theta,99}$ with $\Ray$ for different Prandtl numbers for the lower and the upper upper branch of the roll solution is shown in figure \ref{fig_HeightVsRa}(b).
While for the lower branch of $R_{1}$ the height varies only slightly with $\Ray$, the extension of the  upper branch increases  monotonically with the Rayleigh number.
The rate of increase with $\Ray$ as well as the absolute values of the height depend strongly on the Prandtl number.
For small values of $\Pran$ they are larger than for higher values of $\Pran$.
Specifically, the increase of the height of the upper branch solution with  Rayleigh number follows a linear law (for sufficient distance from the turning point).
The slope increases with decreasing Prandtl number from about $0.4$ for $\Pran=3$ to 8.1 for $\Pran=0.7$.
The results show that for small Prandtl number the roll solutions extend remarkable far above the plate,
and their reach seems to increase with $\Ray$, independent of $\Pran$.
%Furthermore, independent of $\Pran$ the increase seems to be unbounded.

\begin{figure}
\centering
\includegraphics[]{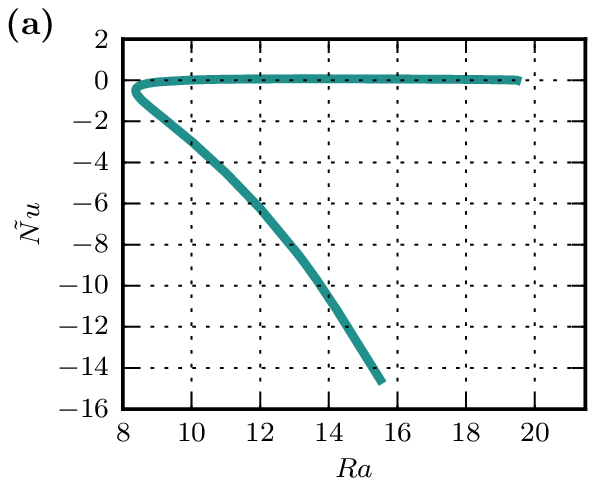}\includegraphics[]{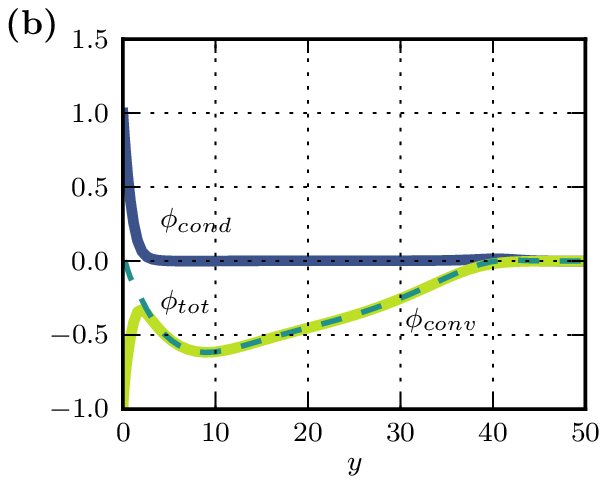}
\caption{%In (a) the 
Nusselt number $\tilde{Nu}$ %is shown 
as a function of the Rayleigh number for the secondary solution $R_{1}$ with $k=k_{c}$ at $Re=0$ is shown in (a).
In panel (b) the contributions of the conductive (blue) and convective (yellow) heat flux are shown as functions 
of the wall-normal coordinate $y$ for the upper branch of $R_{1}$ at $\Ray=15.5$.
The resulting total heat flux is shown by the dashed cyan line. The quantities are normalized with respect to the Rayleigh number.}
\label{fig_Nusselt}
\end{figure}

The secondary roll solutions have different heat transport properties than the base flow.
In figure \ref{fig_Nusselt}(a) the Nusselt number $\tilde{\Nus}$ is shown 
%in dependence on the 
as a function of Rayleigh number for the roll solution $R_{1}$ with $k=k_{c}$ and $\Pran=1$. To rationalize the negative values of $\tilde{\Nus}$ %, which, on first glance, might be confusing, 
the contributions of the convective and the conductive heat transport for a state
on the upper branch of $R_{1}$  are shown as a function of $y$ in figure \ref{fig_Nusselt}(b).
Especially for the upper branch, the heat transport due to convection has large negative values, leading to negative values of the 
integrated quantity $\tilde{\Nus}$ with large modulus. 
The figure shows that for the secondary solution $R_{1}$ the conductive heat transport is only relevant close to the wall and quickly
drops to zero for larger values of $y$. Furthermore, the plot illustrates that indeed the convective heat transport causes the large negative values of $\tilde{\Nus}$ 
(the integral over the dashed curve). 
%The negative convective heat transport can be rationalized as follows:
If only the deviations from the base flow are considered,  
the up-flow and down-flow regions 
in the roll state $R_{1}$
are approximately of equal strength (see figure \ref{fig_FlowFieldsRa13}(b) and (c))
and the  temperatures in both regions are comparable as well. Since the base flow has a negative wall-normal component,
the symmetry is broken and the negative contribution to the convective heat flux due to the down-flow region
becomes larger than the positive contribution of the up-flow region. Thus, the average convective heat flux is negative for $R_{1}$.
However, there are also turbulent states or coherent states with high asymmetry
for which positive values of $\tilde{\Nus}$ are obtained.
%possible in the system, e.g in t

\begin{figure}
\centering
\includegraphics[]{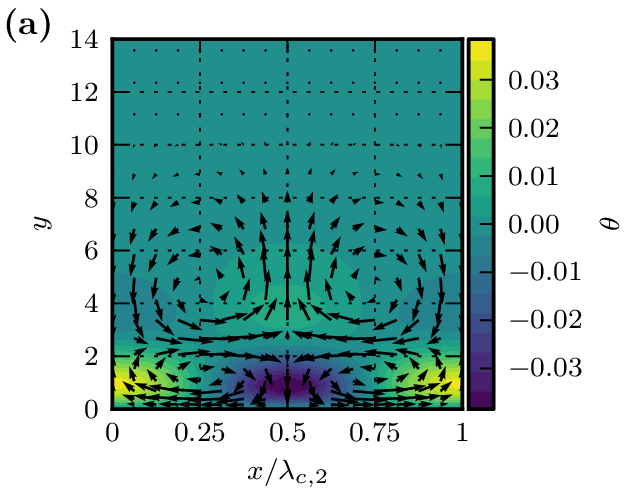}\includegraphics[]{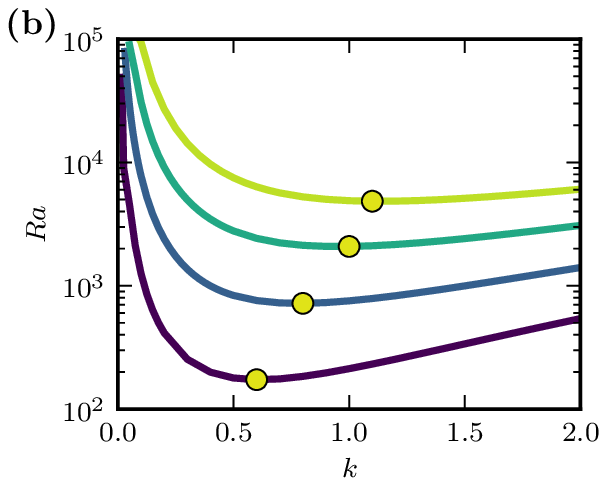}
\caption{In panel (a) a visualization of the exact solution $R_{2}$ that bifurcates from the second two-dimensional instability is shown. The wavenumber is $0.612$ and the Rayleigh number is $167$. 
Panel (b) shows stability curves for the second to fifth two-dimensional unstable modes for $\Pran=1$. 
The critical Rayleigh numbers are marked with a yellow circle.\label{fig_SecUnstableMode}}
\end{figure}

%%%%%%%%%%%%%%%%%%%%%%%%%%%%%%%%
%%%%%%%%%%%%%%%%%%%%%%%%%%%%%%%%
%%%Second Unstable Mode%%%%%%%%%
%%%%%%%%%%%%%%%%%%%%%%%%%%%%%%%%
For high Rayleigh numbers another two-dimensional eigenmode becomes unstable.
For $\Pran=1$, the critical wave number for this mode is $0.612$ and the critical Rayleigh number is $173$. 
This bifurcation is subcritical, and gives rise to an equilibrium solution henceforth referred to as $R_{2}$.
The state $R_{2}$ has a set of four rolls staggered on top of each other, 
with the top pair similar to $R_{1}$ and the bottom one rather flat,
as shown in figure \ref{fig_SecUnstableMode}(b).
The second pair of rolls of the  $R_{2}$ solution  has its centre at a distance of about $5$ from the plate
where the temperature gradient  is already quite low. 
This is reminiscent of the free-stream coherent structures of isothermal ASBL described by \cite{Deguchi2014c}  and \cite{Kreilos2016},
which have vortices at a distance to the plate where the shear gradient is low.

For even higher Rayleigh numbers further modes become unstable. 
Each additional unstable mode has one more layer of rolls. 
The stability curves for the second to fifth two-dimensional unstable modes
are shown in figure \ref{fig_SecUnstableMode}(b).
The critical Rayleigh numbers for the third, fourth and fifth unstable mode are $719.2$, $2082$ and $4846$, respectively.
The corresponding critical wave numbers are $0.78$, $1.0$ and $1.1$.
Since they are very unstable, the ECS created in these bifurcation are of minor importance for the dynamics and thus not treated in this paper. 

\subsection{Stability Analysis for $\Rey\neq0$ and $\Ray\neq0$\label{sec_StabReVar}}
The stability of the base flow for the case that shear and buoyancy are present %
can take advantage of the fact that a generalization of 
Squire's theorem holds for this flow \citep{Gage68}, so that it suffices 
to know the stability boundary for transversal and longitudinal modes to
obtain a complete stability diagram. 

A stability analysis of modes that depend only on the spanwise wavenumber
and are invariant in the streamwise direction (longitudinal modes) yields,
for any Prandtl number, 
the same critical Rayleigh numbers as in the case of vanishing Reynolds number.

\begin{figure}
\centering
\includegraphics[]{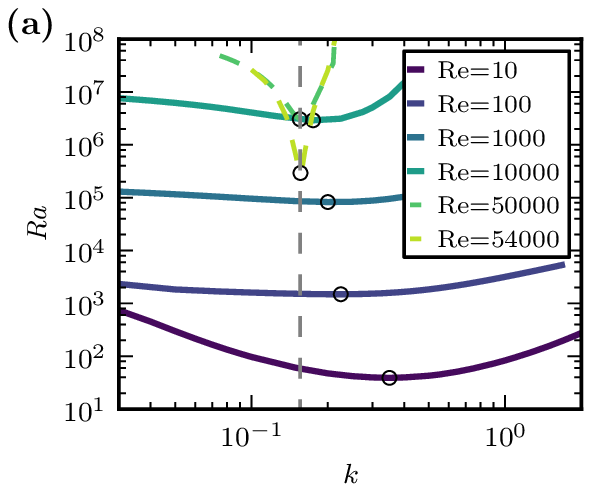}\includegraphics[]{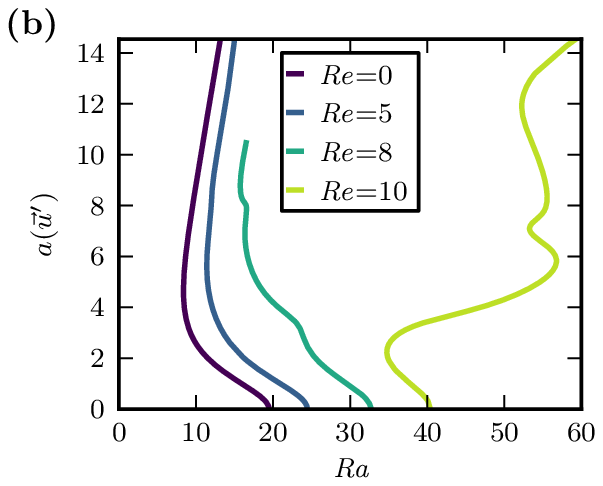}
\caption{Panel (a) shows stability curves of the most unstable transversal mode in dependence on the Reynolds number. For each Reynolds number a black circle marks the critical Rayleigh number. 
The critical wave number of isothermal ASBL is marked by a dashed gray line. In panel (b) bifurcation
diagrams of the two-dimensional transversal waves bifurcating from the laminar flow are shown for various Reynolds numbers.\label{fig_TASBLStabilityReAndRaVar}}
\end{figure}

Initially, for spanwise invariant modes (transversal modes)  an increase of the  Reynolds number has
a stabilizing effect. Figure \ref{fig_TASBLStabilityReAndRaVar}(a) shows stability curves of the most unstable transversal mode for various Reynolds numbers.
Up to a Reynolds number of approximately $35000$ an increase of $\Rey$ results in larger critical 
Rayleigh number. If the Reynolds number is further increased the transversal modes are rapidly destabilized again
until at $Re_{c}=54379$ they are unstable for $\Ray=0$.  The critical wave numbers move monotonically towards the critical wave number for the isothermal case with increasing Reynolds number. 

%For various Reynolds numbers 
We track the travelling waves solutions which bifurcate from the base flow when the transversal mode becomes unstable
for various Reynolds numbers.
The bifurcation diagrams of the transversal waves with $k=k_{c}=0.4304$ %are shown for different Reynolds numbers
in figure \ref{fig_TASBLStabilityReAndRaVar}(b) 
%The results 
show that for finite values of $Re$ the bifurcation is still subcritical in the Rayleigh number, 
but the  turning point of the solution branch moves to higher values of $\Ray$.

\begin{figure}
\centering
\includegraphics[]{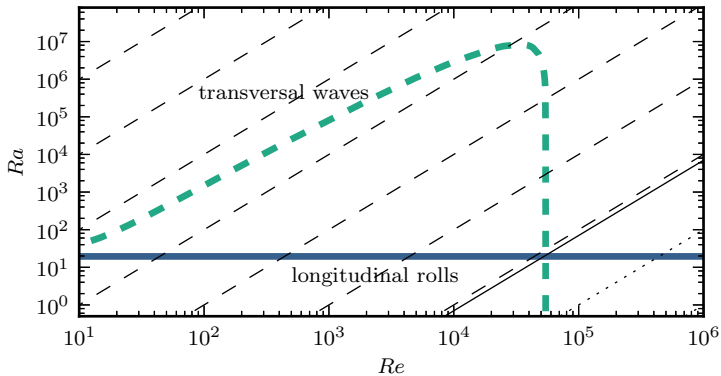}
\caption{Stability diagram of TASBL in dependence on $\Ray$ and $\Rey$. Critical Rayleigh numbers for longitudinal rolls and transversal waves are shown using solid blue and dashed cyan lines.
The solid black line corresponds to the critical Richardson number $\Ric_{c}=7\cdot 10^{-9}$.
% for a change of the 
where the instability switches from longitudinal rolls to transversal waves.
Lines of constant Richardson number are shown using dashed and dotted  lines for $\Ric<\Ric_{c}$ and $\Ric>\Ric_{c}$, respectively.
The values of the Richardson numbers decrease by factors of $10^2$ starting with $10^{4}$ in the upper left corner. 
\label{fig_TASBLStabilityComplete}}
\end{figure}

For systems that are thermally stratified and  under the influence of shear an often used dimensionless number is the Richardson number,
\begin{equation}
\Ric =  \frac{g \beta T_{0} \delta}{U_{\infty}^{2}}=\frac{\Ray}{\Rey^{2}\Pran}, \label{eqn_RichardsonNo}
\end{equation}
which relates the buoyancy to the flow gradient. 
The value of $\Ric$ is a measure for the importance of natural convection compared to forced convection.
%If $\Ric$ is small, effects of natural convection are of minor importance while if $\Ric$ is large, forced convection is negligible. 
As in the study of  \cite{Gage68} for Rayleigh-B\'enard-Poiseuille flow it is possible to give a critical value of 
the Richardson number which marks the transition between natural and forced convection.
%from one kind of instability to the other.
For ASBL over a heated plate we obtain a critical Richardson number of approximately  
$\Ric_{c} = 19.47/(54379^{2} \cdot 1.0) \approx 7\cdot 10^{-9}$.
For smaller Richardson numbers the system has an instability to transversal rolls or waves while
for larger values of $\Ric$ the system has an instability to longitudinal rolls.
The resulting complete stability diagram for the flow (for $\Pran=1$) with the critical Rayleigh numbers of transversal and
longitudinal modes 
%in dependence on 
for different $\Rey$ and lines of constant Richardson number, is shown in figure \ref{fig_TASBLStabilityComplete}.

%%%%%%%%%%%%%%%%%%%%%%%%%%%%%%%%%%%%%%%%%%%%%%%%%%%%%%%%%%%%%%%%%%%%%%%%%%%%%%
%%%%%%%%%%%%%%%%%%%%%%%%%%%%%%%%%%%%%%%%%%%%%%%%%%%%%%%%%%%%%%%%%%%%%%%%%%%%%%%
%%%%%%%%%%%%%%%%%%%%%%%%%%%%%%%%%%%%%%%%%%%%%%%%%%%%%%%%%%%%%%%%%%%%%%%%%%%%%%%

\section{Two-dimensional solutions and subcritical transient chaos
\label{secSubChaos}}
In this section we analyse the stability of the two-dimensional secondary solutions and describe
the bifurcating states. 
The analysis is restricted to the secondary states with a wave number equal to the critical one and
to two-dimensional 
superharmonic disturbances with wave numbers that are integer multiples of the ones of the secondary solution. 
The stability analysis of the solutions is performed using an \textit{Arnoldi}-method included in the Channelflow-package and adapted to work with an additional temperature field. 
The calculations use a computational domain with $L_{x}=0.1$, $L_{z}=\lambda_{0}$ and $L_{y}=40$ and a resolution of $N_{x}\times N_{y} \times N_{z}=4 \times 65 \times 32$, which is a good
approximation of the two-dimensional system.

The stability analysis shows that the lower branch of the secondary solution $R_{1}$ has one unstable direction. 
Thus, it is an edge state \citep{Skufca2006,Schneider2007} for the two-dimensional system because its stable manifold can separate two regions of the state space.
Right after the turning point the upper branch of the solution is stable. 
When the Rayleigh number is increased, a pair of complex conjugate eigenvalues becomes unstable at $\Ray\approx 9.15$, indicating a Hopf bifurcation. This bifurcation creates a stable periodic orbit which we
refer to as $PO_{1}$ in the following. A supplementary movie illustrating the dynamics of the orbit is available online.

\begin{figure}
\centering
\includegraphics[]{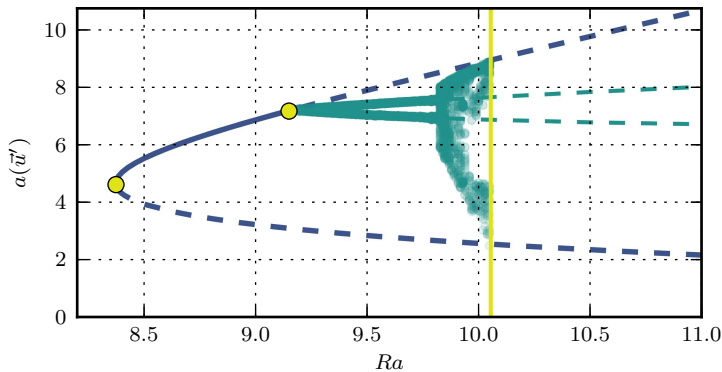}
\caption{Bifurcation diagram of $R_{1}$ with wave number $k=k_{c}$ at $\Pran=1$ and $\Rey=0$.
The curve for $R_{1}$ is shown as a blue line.
Chaotic and periodic states are indicated by cyan points representing minima and maxima of $a(\vec{u}')$ 
in the course of time. In addition the dashed cyan lines show the minimum and maximum of $a(\vec{u}')$ for the orbit $PO_{1}$.
The bifurcation points of $R_{1}$ and $PO_{1}$ are marked by yellow circles. The solid yellow line marks the position of the 
boundary crisis bifurcation near $\Ray=10.0$ which destroys the attractor. \label{fig_Crisis}}
\end{figure}

If the Rayleigh number is further increased, the periodic orbit
undergoes a Neimark-Sacker bifurcation \citep{Kuznetsov1998} which creates a stable torus. 
Further bifurcations then create a chaotic attractor.
A supplementary movie illustrating the dynamics of a trajectory on the chaotic attractor is available online.
The bifurcation diagram, sampled by plotting maxima of $a(\vec{u}')$ for trajectories on the attractor, is shown in figure \ref{fig_Crisis}.
Eventually, the chaotic attractor is converted into a chaotic saddle by  a boundary crisis bifurcation \citep[e.g.][]{Lai2011}.
For $k=k_{c}$ this happens at $\Ray\approx 10.06$. 
Following the boundary crisis, for $\Ray<\Ray_{c}$ initial conditions can be found which transiently show
chaotic dynamics before they finally return to the base flow (supplementary movie online). 
The bifurcation cascade that eventually creates transient chaos in this open flow is akin to those that were found to exist in %the 
internal flows like plane Couette flow \citep{Kreilos2012}, 
pipe flow \citep{Avila2013}, or plane Poiseuille flow \citep{Zammert2015}.  
An important difference is that in the present case the exact solution which is the starting point of the cascade originates from a bifurcation of the base flow. 

\section{Three-dimensional states and localization \label{sec3Dstates}}
Up to this point we have considered two-dimensional exact coherent states, with a modulation in spanwise
or streamwise and normal direction. % of the flow but, of course, the real physical system is three-dimensional. 
Here, we turn to three-dimensional state, in particular states that are either extended or localized in both
spanwise and streamwise direction. 
For all solutions in this section the Prandtl number is fixed at $\Pran=1$ and there is no flow, 
Reynolds number is $\Rey=0$.

\subsection{Straight rolls, squares and hexagons}
\begin{figure}
\centering
\includegraphics[]{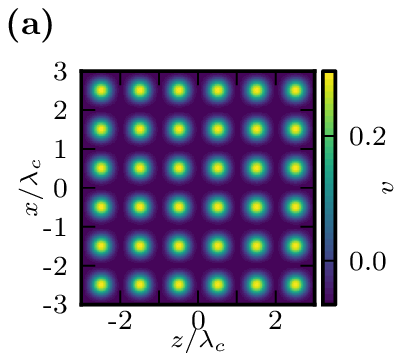}\includegraphics[]{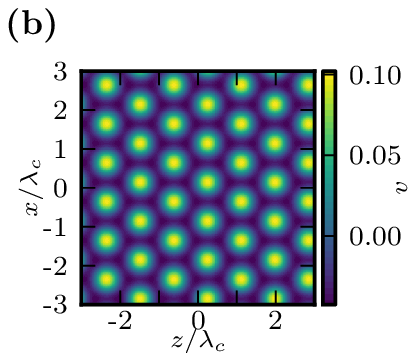}\includegraphics[]{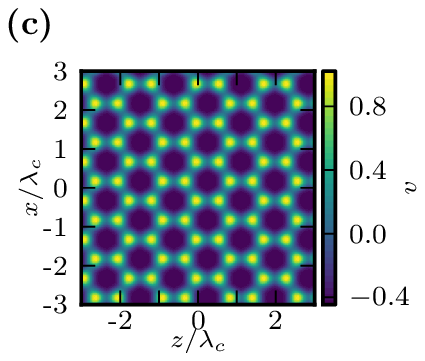}
\caption{Visualization of the three-dimensional secondary solutions at $\Ray=17$. Panel (a) shows the square solution, and panels (b) and (c) 
the hexagonal  solutions  of $l-$ and $g-$type, respectively. For all panels the wall-normal velocity (deviation from laminar profile) 
is shown in a plane parallel to the plate at $y\approx 1.5$.\label{fig_3DPattern}}
\end{figure}

Using a Newton method we can identify different secondary solutions of the three-dimensional system 
which bifurcate from the base profile at the critical Rayleigh number.
The simplest solutions which also exist in the two-dimensional case 
are straight rolls which are translationally invariant along an axis parallel to the plates.
In addition to the straight rolls, a secondary solution exists which shows a square pattern of up-flow regions. 
This solution has  equal wave numbers in both directions parallel to the plate, 
as shown 
in figure \ref{fig_3DPattern}(a).
The visualization shows the wall-normal velocity in a plane parallel to the plate in a distance of $1.5$. 
A continuation of this ECS in Rayleigh number shows that the turning point is
at slightly lower Rayleigh numbers than for the two-dimensional roll-solution $R_{1}$. 
The upper branch of this three-dimensional state again reaches remarkably far above the plates.

It is also possible to identify  two different kinds of secondary solutions with hexagonal pattern.
They are formed by superpositions of three solutions with wave vectors rotated by $2\pi/3$.
%For the hexagonal patterns the wave numbers  in the two directions perpendicular to the plate differ by a factor of $\sqrt{3}$.
One of the hexagonal  patterns has a up-flow region in the centre of each hexagon while for the other one the centre
contains the down-flow region. Following the nomenclature of Rayleigh-B\'{e}nard convection \citep{Getling1998} they
can be named $l-$ and $g-$cells. Visualization of both types of hexagons are shown in figure \ref{fig_3DPattern}(b) and (c).

\begin{figure}
\centering
\includegraphics[]{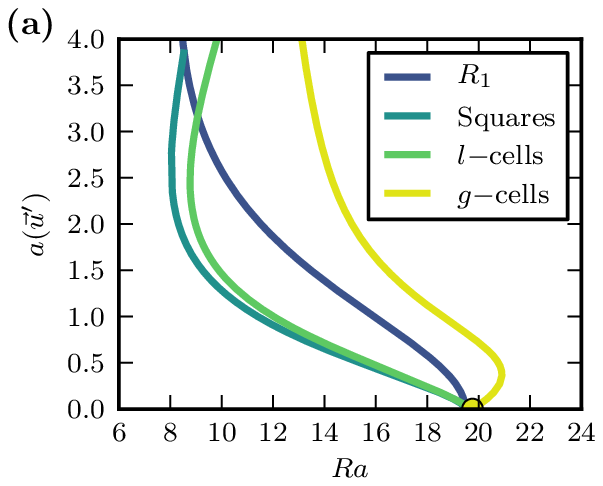}\includegraphics[]{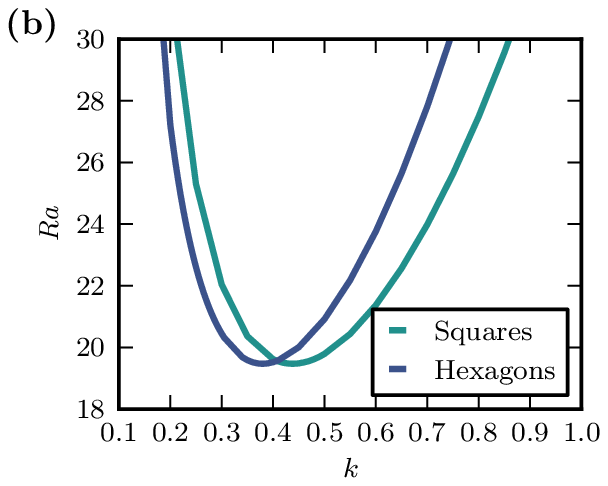}
\caption{Panel (a) shows a bifurcation diagram of the four equilibrium patterns that exist in the three-dimensional system for $\Rey=0$.
For each solution the wave number is $k=k_{c}=0.4305$. In panel (b) the stability lines for 
the hexagonal and the square instability  are compared.\label{fig_BifDiag3D}}
\end{figure}

The bifurcation curves for a wavelength equal to the critical one of the square pattern is shown in figure \ref{fig_BifDiag3D}.
The figure shows that the bifurcation for the $g$-hexagons is initially supercritical and turns to lower $\Ray$ only
with increasing amplitude, while $l$-hexagons
bifurcate subcritically right from the base flow. The lowest values in $\Ray$ are achieved for $l$-hexagons.

For a given wave number the critical Rayleigh numbers for the square pattern and the hexagonal pattern differ slightly.
The difference is too small to be noticeable in figure \ref{fig_BifDiag3D}(a). In figure \ref{fig_BifDiag3D}(b) the dependence of critical Rayleigh number on the wave number 
is shown for both patterns. In the presence of shear both the hexagonal and the square pattern instabilities
move to higher Rayleigh numbers and the two-dimensional longitudinal rolls are the first instabilities to appear.

\subsection{Localized states\label{secLocStates}}

\begin{figure}
\centering
\includegraphics[]{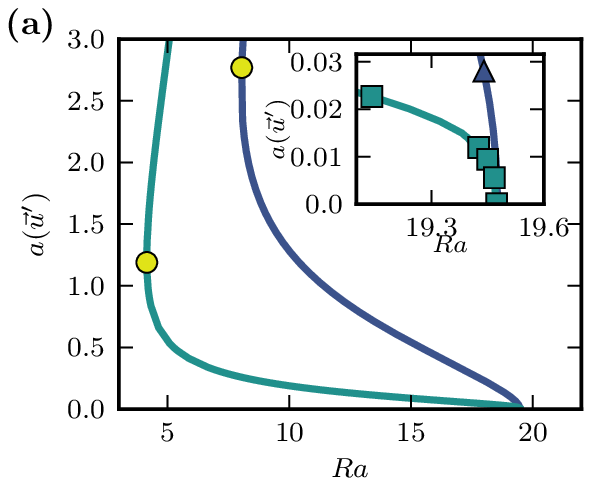}\includegraphics[]{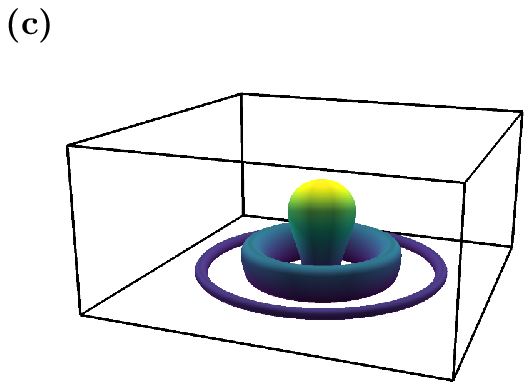}\\
\includegraphics[]{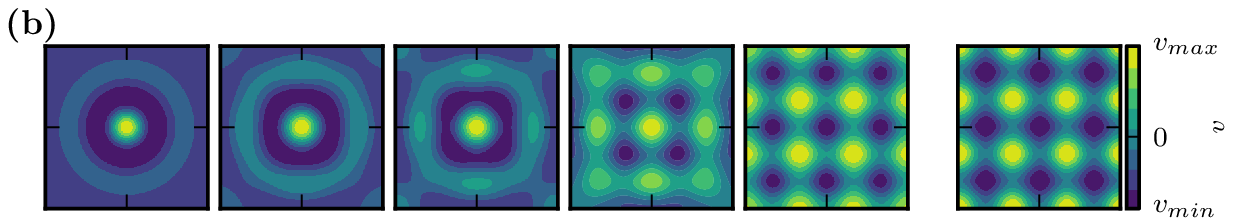}
\caption{Panel (a) shows the bifurcation diagram for the doubly-localized equilibrium solution $C_{1}$ (cyan)  and the spatially
extended equilibrium (blue). The inset shows the bifurcation diagram close to the point where $C_{1}$ connects to the spatially extended pattern. 
In panel (b) the states which are marked with the squares in the inset of (a) are visualized. 
The plots show the wall-normal velocity $v$ in a plane parallel to the plate in a distance of $1.5$. The leftmost picture
corresponds to the square at the lowest Rayleigh number. The rightmost panel in (b) shows the wall-normal velocity $v$ for the state marked in (a) by the triangle. 
In panel (c) the iso-surface $v=0.0005$ (deviation from laminar) is shown for the doubly-localized equilibrium $C_{1}$ at $\Ray=17.0$, $\Rey=0$ and $\Pran=1$. The iso-surface is coloured according to the distance to the
plate from blue to yellow. }
\label{fig_TASBLConvecton}
\end{figure}

\begin{figure}
\centering
\includegraphics[]{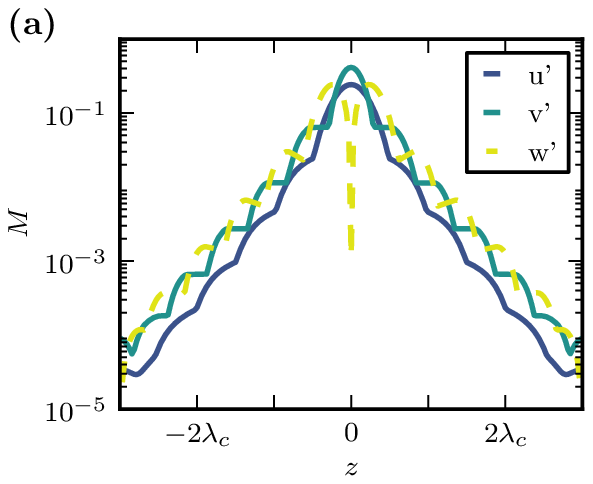}\includegraphics[]{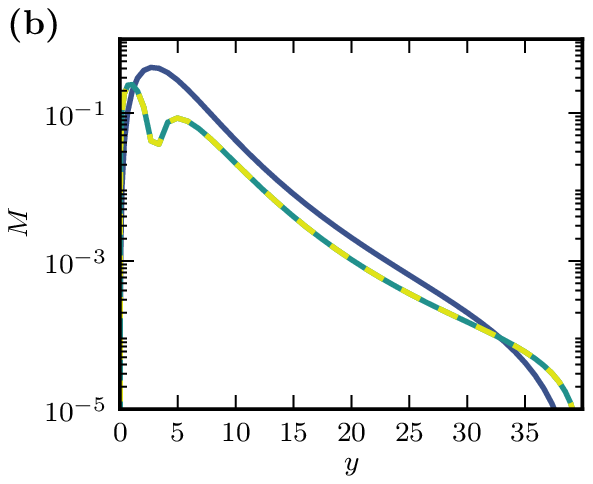}
\caption{Maximal values $M$ of the streamwise (blue), spanwise (yellow) and wall-normal (cyan) velocity
component for the localized equilibrium at $\Ray=17$. 
Panel (a) shows the dependence on the wall-parallel coordinate $z$ and panel (b) the dependence on
the wall-normal coordinate $y$. \label{fig_Linf}}
\end{figure}

With increasing distance to their bifurcation points the different secondary solutions undergo several three-dimensional instabilities. 
An interesting bifurcation of the square solution, which can be found in a domain with $L_{x}=L_{z}=6\lambda_{c}$ and $L_{y}=40$ (resolution
$N_{x}\times N_{y} \times N_{z}=192\times 49 \times 192$), is located slightly below $\Ray_{c}$.
The bifurcation is connected with a long-wavelength instability \citep{Melnikov2014,Chantry2013} in two directions and creates a three-dimensional solution which for lower Rayleigh numbers becomes spatially localized in both directions parallel
to the plate. The bifurcation diagram in figure \ref{fig_TASBLConvecton}(a) shows the spatially extended square pattern together with the bifurcating %tertiary 
solution. 
For the Rayleigh numbers marked in the bifurcation diagram flow visualizations of the %tertiary and the secondary 
solutions are given in figure \ref{fig_TASBLConvecton}(b).
The visualizations illustrate how the spatially extended pattern turns into localized states.
% bifurcates from
Slightly below the bifurcation point, the flow pattern looks like the one of the spatially extended state with an additional modulation.
With decreasing Rayleigh number this modulation quickly becomes larger until a localized and rotational symmetric state is reached.

The three-dimensional visualization of the state at $\Ray=17$ which is shown in  figure \ref{fig_TASBLConvecton}(c) impressively illustrates the localization and the symmetry of the state. 
For localized convection states akin to the identified ECS but only localized in one spatial direction the term \textit{convecton} has been 
introduced \citep[e.g.][]{Blanchflower99,LoJacono2011,Beaume2013}.
Adopting this notation, we will call the doubly-localized equilibrium a convecton and  use the abbreviation $C_{1}$. 

As a measure for the localization of the state $C_{1}$ we use the maximum-norm, 
\begin{equation}
%\mathcal{L}^{\infty}(u')=\max_{y,x}|u(x,y,z)'|,
M(u')=\max_{y,x}|u(x,y,z)'|,
\end{equation}
here given for the streamwise component $u$ and in dependence of the spanwise coordinate. 
In figure \ref{fig_Linf} the maximum-norm of all velocity components in %dependence on 
the wall-parallel (a) and wall-normal (b) directions are shown. 
Since the localization is quite strong, the state exists also in larger domains and is not an artefact of the boundary conditions.
In the directions parallel to the plate and also in wall-normal direction, the decay of the velocities is in good agreement 
with an exponential decay. 
In contrast to the doubly localized exact solution found in plane Poiseuille flow \citep{Zammert2014b}, the decay does not become 
algebraic for large distances. The difference can be rationalized by the fact that in plane Poiseuille flow the  shear creates a 
large-scale flow around the exact solution which decays algebraically. 
Since in the present case the Reynolds number is zero, there is no shear and no large-scale flow. 
Thus, the decay is exponential over a large range.

The turning point for $C_{1}$ lies at $\Ray \approx 4.14$. Close to the turning point,
the localization becomes weaker again and especially  
the upper branch becomes very wide and very difficult to track numerically.

\begin{figure}
\centering
\includegraphics[]{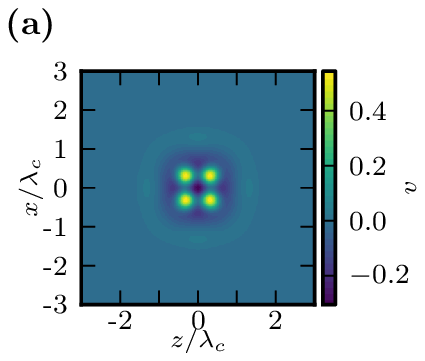}\includegraphics[]{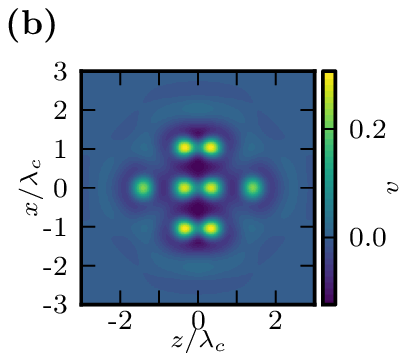}\includegraphics[]{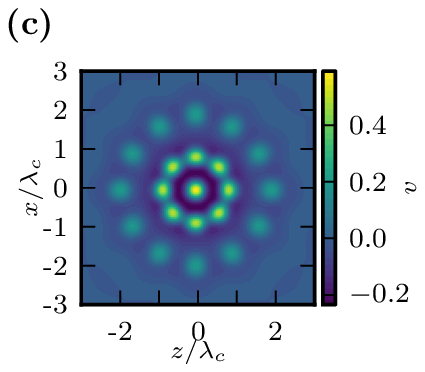}
\\
\includegraphics[]{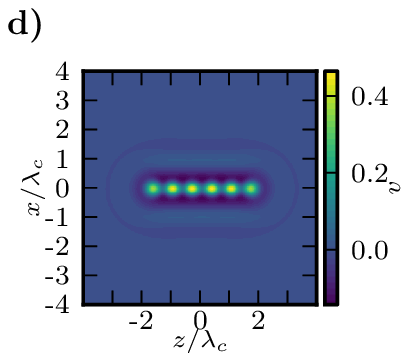}\includegraphics[]{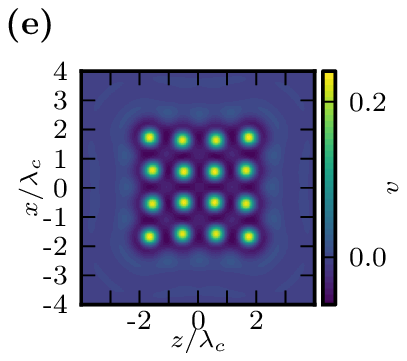}\includegraphics[]{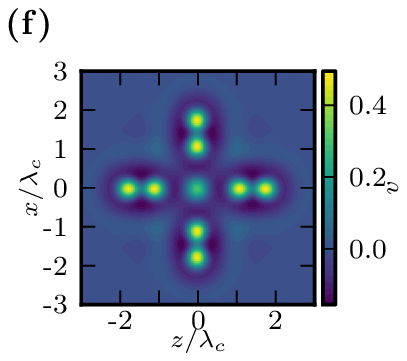}
\includegraphics[]{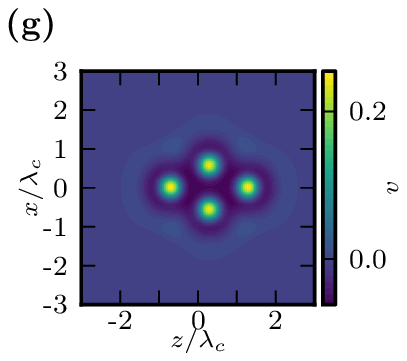}\includegraphics[]{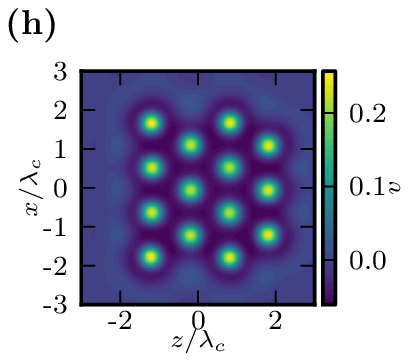}\includegraphics[]{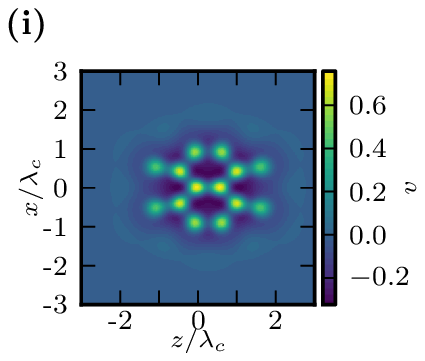}

\caption{Different doubly-localized solutions at $\Ray=17.8$. The Prandtl number is 1 and the Reynolds number is 0.
The pictures show the wall normal velocity in a plane parallel to the lower plate at $y\approx 1.5$.\label{fig_fullLocSolutions}}
\end{figure}

In addition to the localized equilibrium $C_{1}$, many other localized states can be found. 
Starting with the spatially extended secondary solutions, we use two-dimensional window functions \citep{Gibson2014,Brand2014} to create initial guesses for Newton-searches for localized equilibrium solutions.
A two dimensional window function $W(x,z)$ can be written as
a product of two one-dimensional window functions.
We use two different types of one-dimensional window functions, a Gaussian
\begin{equation}
W_{1}(x)=A\exp{\left(-B x^{2}\right)} \label{Window2}
\end{equation}
and the function
\begin{equation}
W_{2}(x)=\frac{1}{4}\left(1 + \tanh{\left( \frac{6(A-x)}{B} +3 \right)} \right)\left(1 + \tanh{\left( \frac{6(A+x)}{B} +3 \right)} \right). \label{Window1}
\end{equation}
introduced by \cite{Gibson2014}. For appropriately chosen parameters $A$ and $B$
the obtained initial conditions are good initial guesses for equilibrium solutions. In figure \ref{fig_fullLocSolutions}
visualizations of different localized equilibrium solutions that were found for $\Ray=17.8$ and $\Pr=1$ are shown.
The states shown in (a)-(c) were obtained from initial conditions created using the window function $W_{1}$ and for the state in (d) and (e)
the window function $W_{2}$ was used. The equilibrium shown in (f) was obtained from an initial guess that was
constructed by combining flow fields of several $C_{1}$ solutions. The state shown in (g)-(i) are obtained by applying window functions to the hexagonal patterns.

\section{Summary and outlook\label{SecConcl}}
The present study extends the family of sheared convection problems from 
planar sheared convection \citep{Clever1992} and Poiseuille-Rayleigh-B\'enard flow \citep{Nicolas2002}
to the case of the ASBL, as a step towards analyzing developing boundary layers.
In all flows the transition is connected with roll states that are preferentially aligned with the 
flow direction, at least for Reynolds numbers close to the threshold.  For higher 
Reynolds numbers, transverse rolls are preferred. 
The critical Rayleigh number shows a strong dependence on Prandtl number, 
which can be rationalized by relating it to 
an effective Rayleigh number that takes the temperature difference across the velocity field
and the thickness of the thermal boundary layer into account. 

The presence of a subcritical instability gives rise to a wide variety of secondary states, in particular
of localized solutions of various shapes. While some of these states are reminiscent 
of plumes and other structures observed in thermal boundary layers, further studies
of their dynamical properties will be required in order to establish this link more formally.
It will be also be interesting to see how a spatial development as in Blasius boundary layers
will influence the structures and their properties. Eventually, the study of 
the stability of convection combined with suitable boundary layer flows should provide
a more reliable description of the transition from laminar to turbulent boundary layers
and thus of the transition to the ultimate regime in thermal convection, 
\citep[]{Grossmann2002}.

Finally, it is intriguing that the upper branches of the coherent structures reach far into the 
free-stream region above the boundary layers. Certain ECS in shear boundary layers show 
similar properties \citet{Deguchi2014c,Kreilos2016}. The observation that
the turbulent ASBL has remarkably thick boundary layers 
\citep{Schlatter2011,Bobke2016} thus suggests that a study of the thermally heated ASBL in conjunction
with investigations of the various exact coherent structures could also shed light on the
large scale properties and organization of the flow beyond the transitional regime discussed here.

\section*{Acknowledgements} We thank John  F. Gibson for providing and maintaining {\it channelflow}.
This work was supported in parts by the Deutsche Forschungsgemeinschaft (DFG) within FOR 1182 
and by Stichting voor Fundamenteel Onderzoek der Materie (FOM) within the program "Towards ultimate turbulence".

\appendix

\section{Direct numerical simulation\label{AppendixDNS}}
For the study of ASBL with heated plate %in  this paper 
the Channelflow-code \citep{J.F.Gibson2012} was adapted to integrate the 
set of coupled equations (\ref{OBeq1}-\ref{OBeq3}).
The velocity field is calculated as in the isothermal case, and the
coupling of the temperature field on the velocity field  is incorporated in the nonlinear part that
is treated explicitly in the code.

The time integration of the temperature field is analogous to the velocity field. 
The total  velocity field as well as the total temperature $\theta$ can be written as a sum
of the laminar or base profile and deviations,
\begin{equation}
 \theta(\vec{x})=\theta(\vec{x})'+T(y)
\end{equation}
\begin{equation}
 \vec{u}(\vec{x})=\vec{u}(\vec{x})'+\vec{U} 
\end{equation}
%With We plug  this into 
Equation %\ref{TempEq} 
\ref{OBeq2} then gives
\begin{equation}
 \partial_{t} \theta' + \underbrace{(\vec{u} \cdot \nabla) \theta}_{A(\vec{u},\theta)}= \underbrace{\kappa \Delta \theta'}_{D\theta} +  \underbrace{\kappa \partial^{2}_{y} T(y)}_{C} \label{TempEq2} 
\end{equation}
The time integration takes place in the Fourier-space, where %. Therefore, it is necessary to Fourier transform equation 
the different terms in %\ref{TempEq2} 
\ref{OBeq2} become, symbolically,
\begin{equation}
  \partial_{t} \tilde{\theta}  + \tilde{A}(\vec{u},\theta)= \tilde{D}\tilde{\theta} + \tilde{C}  \label{TempEq3} 
\end{equation}

Channelflow offers different time-integration schemes
which were all adapted to work with the additional temperature field.
Using %Analogous to 
the isothermal case described in the Channelflow-manual as an example, we here 
explain the time-stepping of the temperature field for the Crank-Nicolson/Adams-Bashforth scheme (CNAB2).
Let $\tilde{\theta}^{n+1}$ denote the approximation to $\tilde{\theta}$ at time $n\Delta t$.
Then, %Therefore, 
we first approximate the terms of equation \ref{TempEq3} as follows:
\begin{equation}
  \partial_{t} \tilde{\theta}^{n + 1/2}= \frac{\tilde{\theta}^{n+1}-\tilde{\theta}^{n}}{2} + \mathcal{O}(\Delta t)
\end{equation}

\begin{equation}
  \tilde{D}\tilde{\theta}^{n + 1/2}= \frac{1}{2}\tilde{D} \tilde{\theta}^{n+1}+\frac{1}{2}\tilde{D} \tilde{\theta}^{n} + \mathcal{O}(\Delta t)
\end{equation}

\begin{equation}
  \tilde{A}^{n + 1/2}= \frac{3}{2}\tilde{A}^{n}-\frac{1}{2}\tilde{A}^{n-1} + \mathcal{O}(\Delta t)
\end{equation}

\begin{equation}
  \tilde{C}\tilde{\theta}^{n + 1/2}= \frac{1}{2}\tilde{C}^{n}+\frac{1}{2}\tilde{C}^{n-1} + \mathcal{O}(\Delta t)
\end{equation}
%The approximation of $D$ and $A$ are called Crank-Nicolson and Adams-Bashforth, respectively.
Substituting these approximations in equation \ref{TempEq3} gives %we obtain:
\begin{equation}
 \left[ \frac{1}{\Delta t}- \frac{1}{2}\tilde{D} \right] \tilde{\theta}^{n+1}= \left[ \frac{1}{\Delta t}+\frac{1}{2}\tilde{D} \right]\tilde{\theta}^{n}+\frac{3}{2}\tilde{A}^{n}-\frac{2}{2}\tilde{A}^{n-1} \label{TempEq4}
\end{equation}
The right hand side of this equation can be calculated directly since only previous time steps contribute to it.
Equation \ref{TempEq4} has the form of a Helmholtz-equation, 
\begin{equation}
\partial_{y}^{2}\tilde{\theta}^{n+1} - \lambda \tilde{\theta}^{n+1} = R  , \label{HemholtzEqn}
\end{equation}
where
\begin{equation}
\lambda=\frac{1}{\Delta t} +4 \pi^{2} \nu \left(\frac{k_{x}^{2}}{L_{x}^{2}}+ \frac{k_{z}^{2}}{L_{z}^{2}}\right)
\end{equation}
\begin{equation}
R=\left[ \frac{1}{\Delta t }+ \frac{1}{2}D \right] \tilde{\theta}^{n}+\frac{3}{2} A^{n}- \frac{1}{2}A^{n-1} +\frac{1}{2}C^{n+1} +\frac{1}{2}C^{n} .
\end{equation}
Routines for solving these equation are included in the Channelflow-package since similar equations
have to be solved for the velocity fields, too.

In addition to CNAB2 the other time-steppers of Channelflow were adapted to handle an additional temperature field.
In all cases, within one substep, the temperature is integrated forward first. For the calculation of $A(\vec{u},\theta)$ the
velocity field of the previous substep is used. Afterwards, the velocity field is integrated using a 
combination $(\theta^{n+1}+\theta^{n})/2$ of the new and the old temperature fields. 
For the simulation of TASBL presented in this paper we use channeflows SBDF3 integrator
but double-check some simulations using CNAB2.

\begin{figure}
\centering
\includegraphics[]{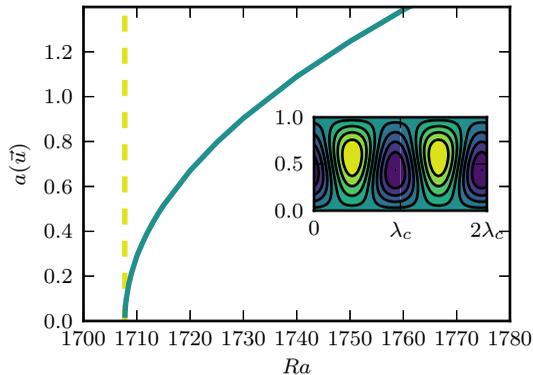}
\caption{Bifurcation diagram of the Rayleigh-B\'enard instability.\label{fig_RayBen} The dashed yellow line marks the position for
the theoretical critical Rayleigh number and the solid line shows the amplitude $a(\vec{u})$ of the convection
solution. The inset shows the temperature field (deviation from conduction state) of the convection solutions at $Ra=1880$. }    % Give a unique label
\end{figure}

To test our implementation several qualitative and quantitative test are performed. 
We calculated the critical Reynolds number of Rayleigh-B\'enard-convection and compared our result to the
known literature value $Ra=1707.76$ \citep{Busse1978}. A calculated bifurcation diagram for the critical wavenumber is shown in figure \ref{fig_RayBen}.
The relative error in critical Rayleigh number obtained with the
code compared to the literature value is of order $10^{-5}$.
As a further (qualitative) test we considered sheared convection. Following the work of \cite{Clever1992}, 
the ECS in isothermal plane Couette flow was continued in Rayleigh number and shown to connect to the roll-solution 
created by the thermal instability. 

\end{document}